\documentclass{osa-article}

\journal{osajournal}

\setprjcopyright

\articletype{Research Article}

\usepackage{lineno}
\usepackage{hyperref}

\begin{document}

\title{Generation of quantum-certified random numbers using on-chip path-entangled single photons from an LED}

\author{Nicolò Leone\authormark{1,*}, Stefano Azzini\authormark{1},  Sonia Mazzucchi\authormark{2}, Valter Moretti \authormark{2}, Matteo Sanna \authormark{1}, Massimo Borghi \authormark{3}, Gioele Piccoli \authormark{4}, Martino Bernard \authormark{4}, Mher Ghulinyan \authormark{4} and Lorenzo Pavesi\authormark{1}}

\address{\authormark{1}Department of Physics, University of Trento, via Sommarive 14, Trento (Italy)\\
\authormark{2}Department of Mathematics and TIFPA, University of Trento, via Sommarive 14, Trento (Italy)\\
\authormark{3}Department of Physics, University of Pavia, via Agostino Bassi 6, 27100, Pavia (Italy)\\
\authormark{4}Centre for Sensors and Devices, Fondazione Bruno Kessler, via Sommarive 18, Trento (Italy)}

\email{\authormark{*}nicolo.leone@unitn.it} 



\begin{abstract}
Single-photon entanglement is a peculiar type of entanglement in which two or more degrees of freedom of a single photon are correlated quantum-mechanically. Here, we demonstrate a photonic integrated chip (PIC) able to generate and manipulate single-photon path-entangled states, using a commercial red LED as light source. A Bell test, in the Clauser, Horne, Shimony and Holt (CHSH) form, is performed to confirm the presence of  entanglement, resulting in a maximum value of the CHSH correlation parameter equal to $2.605 \pm 0.004$. This allows us to use it as an integrated semi-device independent quantum random number generator able to produce certified random numbers. The certification scheme is based on a Bell's inequality violation and on a partial characterization of the experimental setup, without the need of introducing any further assumptions either on the input state or on the particular form of the measurement observables. In the end a min-entropy of $33\%$ is demonstrated.
\end{abstract}

\section{Introduction}\label{section:intro}

Entanglement is one of the most striking feature of quantum physics. The non-classical correlations that entanglement induces in quantum states have been debated since 1935, when Einsten, Podolski and Rosen pointed out what would have been named as the EPR paradox after them ~\cite{Einstein35}. It was only several years later that a way out of the impasse was proposed by John Bell. Its famous inequality~\cite{bell1964} provides indeed a quantitative solution to effectively demonstrate that correlations induced by entanglement cannot be explained classically using any realistic local-hidden variable. Today entanglement represents a resource in many quantum applications, especially in quantum computing and communications~\cite{nielsen2002quantum,Macchiavello04,Bennett92,Ekert91,Ekert98,Pironio10}. However, its exploitation is mainly limited to research laboratories only, still far from being used in real-life devices. This is mainly due to technological complexities related to its generation and management.

In photonics, entangled photons pairs are typically obtained exploiting non-linear optical processes as spontaneous parametric down conversion~\cite{Magnitskiy15} or four wave mixing~\cite{Takesue04}, using suitable laser sources.
On the contrary, single-photon entanglement (SPE)~\cite{Azzini2020} can be more easily generated using only linear optical components and cheap light sources, like LEDs~\cite{Pasini20}. SPE corresponds to quantum correlations between two or more degrees of freedom (DoFs) of a single photon. An example of the possible DoFs used are momentum and polarization\cite{Michler2000,Gadway2009,Pasini20,Leone2022}.
From the mathematical point of view, SPE is totally analogous to the entanglement of two photons, or inter-photon entanglement. In both cases, the Hilbert space is determined by the tensor product of two independent Hilbert spaces: in the case of SPE, these are the spaces associated with the two independent DoFs chosen, while, in the case of the inter-photon entanglement, the two spaces are each one associated to one of the two photons. 
It is from the physical point of view that the differences are more remarkable. Inter-photon entanglement exhibits a non-local phenomenology, while SPE concerns contextual, but local, observations. The meaning of the violation of the Bell inequality is also different in the two scenarios. In the case of inter-photon entanglement, it means that no realistic local hidden variable theory is able to explain the experiment's results, while, in the case of SPE, it means that no realistic non-contextual hidden variable theory is capable of predicting the results. The fact that SPE is a local phenomenon implies that it cannot be used as a substitute of inter-photon entanglement in many quantum applications. However, there can be specific cases in which, thanks to the easier generation and management aspects, it can be exploited for developing entanglement-based applications with potentially larger diffusion. One of them concerns random number generation.

Random numbers are fundamental resources in many different applications, like gambling and gaming, lotteries, computer simulation and cryptography. In cryptography, in particular, the unpredictability of the random sequence ensures the reliability of the encryption protocols. To generate random numbers, random number generators (RNGs) are usually exploited.
Among all RNGs, quantum random number generators (QRNGs)\cite{Collantes16,Ma2016} are the only ones that can generate truly random numbers. Indeed, pseudo-random number generators (PRNGs)\cite{james2020} and true-random number generators (TRNGs)\cite{zhun2001,hu2009} exploit algorithms and noisy/chaotic processes, respectively, to generate random sequences. While the first ones are not truly random by definition, the second ones involve complex physical processes making randomness certification quite hard to obtain. 
On the contrary, for QRNGs such a certification is usually easier and can be done even considering malicious and error-prone implementation or an eavesdropper that is attacking the generator. Considering the level of security, QRNGs can be divided in three categories: device-dependent QRNGs (DD-QRNGs), device-independent QRNGs (DI-QRNGs) and semi-device-independent QRNGs (SDI-QRNGs)\cite{Collantes16}. DD-QRNGs are the less secure, as their randomness certification scheme is based on the perfect characterization of their physical implementation. In perfect conditions, these QRNGs work quite well, but they are unable to cope with any change in their performances, which can possible alter the produced randomness.
DI-QRNsG are instead the most secure QRNGs, since their a-priori randomness certification is independent by the actual characterization of the physical system involved. However, their physical implementations are still particularly complex and challenging, strongly limiting so far their exploitation to research labs only. The SDI-QRNGs represent a trade-off between the easiness of implementation of the DD-QRNGs and the security of the DI-QRNGs: only limited parts of the physical implementation are characterized, treating the others as black boxes.

In this work we propose a photonic integrated circuit (PIC) able to generate and manipulate single photon path entangled states by using typical integrated photonic devices like multi-mode interferometers (MMIs), thermal phase shifters (PSs), Mach-Zehnder interferometers (MZIs) and crossings (CRs)\cite{vivien2016}. The qubit encoding is done using four waveguides: depending on which waveguide the photon is injected, a certain state is codified ~\cite{silverstone2014,wang2018}. The presence of entanglement is validated performing a Bell test\cite{scarani2019}. SPE and Bell inequality violation are used to lower-bound the conditional min-entropy of the generated sequence of measurement outcomes. The min-entropy represents the figure of merit to measure randomness~\cite{Koning09}, and its correct estimation is fundamental for each RNGs to obtain, from the raw sequence of generated bits, a sequence of uniform random digits using the randomness extraction procedure~\cite{Nisan99}.
The reported PIC implements a SDI-QRNG.
The necessary assumptions over which it is based are the use of trusted and characterized detectors and the not-maliciousness of the experimental setup used, which can be, however, considered error-prone.
With respect to a previous work from a few of us~\cite{Leone2022}, the SDI-QRNG here proposed is a PIC exploiting an external commercial red LED as a light source and achieving a high value of min-entropy in the most general and secure scenario one can envision, thus making a significant step towards real-world applications.
 
The paper is organized as follows. In Section \ref{section:chipstructure}, we introduce SPE in the case of path entanglement and we detail the structure of the PIC. Then, in Section~\ref{section:non-ideal}, all the non-idealities of the experiment are considered and their effect on the Bell inequality is taken into account. In Section~\ref{sec:expdemonstration}, we present the experimental data that certify the generation of path-entangled states. In Section~\ref{section:protocol}, we present the experimental data demonstrating our SDI-QRNG based on SPE states of path. Finally, in Section \ref{section:conclusion} we draw the conclusion, and in Section \ref{section:methods} we report the experimental methods.

\section{Single-photon path-entangled states on a PIC}\label{section:chipstructure}
\begin{figure}[h!]
\centering
\includegraphics{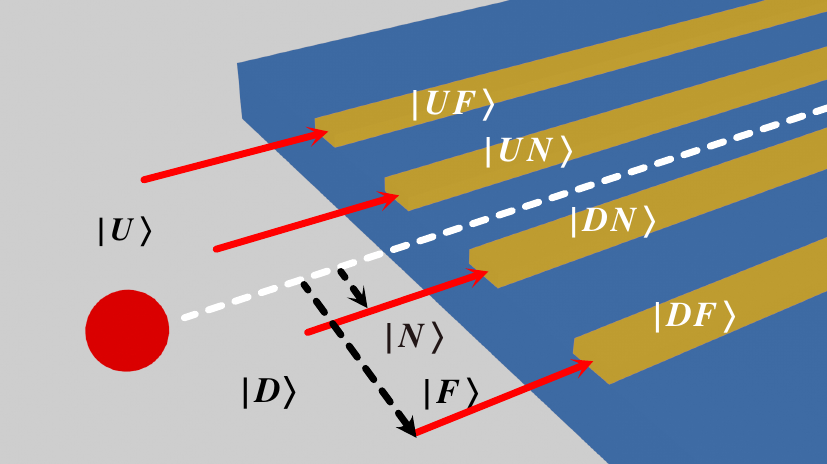}
\caption{Qubit encoding. Two qubits describe the system and are encoded according to the absolute and relative positions of the waveguide in which the photon is injected with respect to the dashed white line. The values of the two qubits are fixed using the following bases: absolute-position (up $|U\rangle$ and down $|D\rangle$) and relative-position (far $|F\rangle$ and near $|N\rangle$).}
\label{fig:qubitencoding}
\end{figure}

\begin{figure}[h!]
\centering
\includegraphics{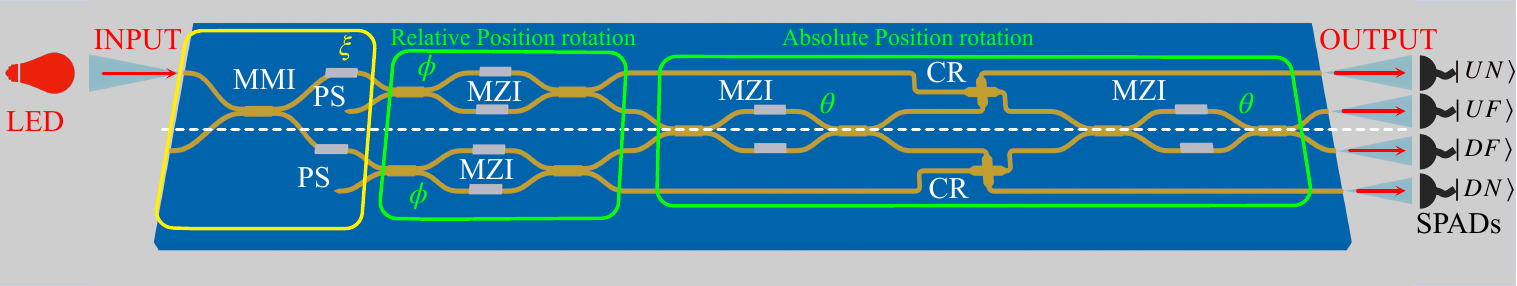}
\caption{Schematic representation of the PIC used for random number generation based on SPE. In cyan, the optical waveguides; in blue, the oxide cladding. A red LED is used as a light source. Light coupling in and out of the chip is performed using tapered optical ﬁbers (transparent cones in the drawing). The PIC can be divided in three parts: generation, relative-position rotation and absolute-position rotation. The generation stage is enclosed by the yellow rectangle on the left side. Here, the entangled state is created. The relative-position rotation corresponds to the first green rectangle from the left: here two MZIs rotate the qubit of relative-position by an angle $\phi$. The absolute-position rotation stage is found in the large green rectangle on the right side: here two MZIs rotate the qubit of absolute-position by an angle $\theta$. At the output, the rotated state is projected onto one of the four states composing the basis of the four-dimensional Hilbert space: $|UF\rangle,|UN\rangle,|DF\rangle$ and $|DN\rangle$. List of abbreviations: MMI, multi mode interferometer; PS, phase shifter; MZI, Mach-Zehnder interferometer; CR, crossing; SPADs, single photon avalanche diodes.}
\label{fig:chipscheme}
\end{figure}

Single-photon entangled states are those states in which at least two DoFs of a single photon are quantum correlated. Here we generate single photon path entangled states by considering four waveguides and two effective DoFs, namely the absolute ($|U\rangle,|D\rangle$) and the relative ($|F\rangle,|N\rangle$) positions of each waveguide with respect to the symmetry axis of the system (dashed white line in Fig.~\ref{fig:qubitencoding}). 
According to the scheme of Fig.~\ref{fig:qubitencoding}, such an Hilbert space can be seen as a $\mathscr{H}=\mathbb{C}^2\otimes \mathbb{C}^2$.
The four states of the Bell basis in such qubit encodig are:
\begin{align}
    &|\phi^{\pm}\rangle=\frac{1}{\sqrt{2}}\left(|UF\rangle\pm|DN\rangle\right),\\
    &|\psi^{\pm}\rangle=\frac{1}{\sqrt{2}}\left(|UN\rangle\pm|DF\rangle\right).
\end{align}
In this work, we focus on the state $|\phi^+\rangle$.
To generate such state, we have designed and fabricated a PIC (a scheme is reported in Fig.~\ref{fig:chipscheme}) based on Silicon Oxynitride ($\text{SiO}_x\text{N}_y$) material, a low-index contrast photonic platform~\cite{Piccoli22}. The structure of the PIC is simple and composed only by linear integrated optical elements: MMIs, PSs, MZIs and CRs. The first part of the PIC (yellow box in Fig.~\ref{fig:chipscheme}) is responsible for generating the SPE state $|\phi^+\rangle$ by exploiting a 50:50 MMI and two PSs setting the relative phase $\xi$ of the entangled state, that, apart from a global non-influent phase, can be written as:
\begin{equation}\label{eq:inputstate}
	|\psi\rangle=\frac{1}{\sqrt{2}}\left(|UF\rangle+ie^{i\xi}|DN\rangle\right),
\end{equation}
with $\xi=\xi_1-\xi_2$, where $\xi_{1(2)}$ is the phase induced by the phase shifter applied to $|UF\rangle(|DN\rangle)$ in the generation stage. By tuning $\xi$, it is eventually possible to precisely obtain the Bell state $|\phi^+\rangle$.
To demonstrate the presence of entanglement, a Bell test\cite{Bell74} in the Clauser, Horne, Shimony and Holt (CHSH) form\cite{Clauser69} is operated. 
Its aim is to quantify the presence of correlations between suitable measurements performed on the two qubits considered. 
To run a Bell test, it is necessary to define four measurement operations to be performed on the entangled state and connected to two observables, $\{A(x)\}_{x=0,1}$ and $\{B(y)\}_{y=0,1}$, each one dependent on a binary variable, $x$ and $y$ respectively. Such measurements must have binary outputs, $a$ and $b$, which can then assume values $\{\pm1\}$. In particular, $A(x)$ is an observable that have to be measured only on one qubit, while $B(y)$ only on the other.
By defining the correlation coefficients $\mathbb{E}(x,y)$ as:
\begin{equation}\label{eq:corrcoeff}
    \mathbb{E}(x,y)=\mathbb{P}(a=b|x,y)-\mathbb{P}(a\neq b|x,y),
\end{equation}
where $\mathbb{P}(a=b|x,y)$ is the conditional probability of observing $(a=b)$ and $\mathbb{P}(a\neq b|x,y)$ is the conditional probability of observing $(a\neq b)$, it is possible to define the correlation function $\chi$ as:
\begin{equation}
    \chi=\sum (-1)^{xy}\mathbb{E}(x,y).
\end{equation}
If $|\chi|>2$, then we are dealing with an entangled state. More precisely, entanglement is a necessary but non sufficient condition for a state to satisfy such an inequality, while separable states always result in $|\chi|<2$\cite{Clauser69}. In the PIC, the Bell test is performed by using the other two stages. The two measurement operations $\{A(x)\}_{x=0,1}$ and $\{B(y)\}_{y=0,1}$ are implemented by a combination of four MZIs with the help of two CRs and four single-photon avalanche diodes (SPADs) that are off-chip. In particular, two MZIs work in parallel to rotate the relative-position qubit by an angle $\phi$ (green box at the center of Fig.~\ref{fig:chipscheme}), while two cascaded MZIs, separetd by a pair of CRs, rotate the absolute-position qubit by an angle $\theta$ (green box on the right side of Fig.~\ref{fig:chipscheme}), where $\phi$ and $\theta$ correspond to $y$ and $x$ in Eq.\ref{eq:corrcoeff}, respectively. A fair implementation of such rotations requires that the same rotation angle is set for both the MZIs relative to the same qubit. 
Then, the rotated SPE state is projected over the four states  $|UF\rangle,|UN\rangle,|DF\rangle$ and $|DN\rangle$ composing the Hilbert space, by means of four waveguides coupled to four off-chip SPADs. 
According to Eq. \ref{eq:corrcoeff}, the correlation coefficients $ \mathbb{E}(\phi,\theta)$ can be written as:
\begin{equation}
    \mathbb{E}(\phi,\theta)=\mathbb{P}(|UF\rangle|\phi,\theta)+\mathbb{P}(|DN\rangle|\phi,\theta)-\mathbb{P}(|UN\rangle|\phi,\theta)-\mathbb{P}(|DF\rangle|\phi,\theta),
\end{equation}
where 
\begin{equation}\label{eq:probrot}
    \mathbb{P}(|ab\rangle|\phi,\theta)=\operatorname{Tr}\left[U(\phi, \theta)\rho U(\phi,\theta)^\dag P_{a}\otimes P_{b}\right],
\end{equation} 
is the probability of the rotated state $U(\phi, \theta)\rho U(\phi, \theta)^\dag$ to collapse on the state $|ab\rangle$ given the couple of angles $(\phi,\theta)$.
It is important to underline that Eq.~\ref{eq:probrot} is essentially equivalent to:
\begin{equation}
    \mathbb{P}(|ab\rangle|\phi,\theta)=\operatorname{Tr}\left[\rho P^{\theta}_a\otimes P^{\phi}_b\right],
\end{equation}
where $\{P^{\theta}_a\otimes P^{\phi}_b\}$ is the set of projection-valued measure (PVM) associated to the observables $A(\theta)$ and $B(\phi)$.
The theoretical form of the correlation coefficients $\mathbb{E}$ and correlation function $\chi$ can be calculated introducing the matrix representation of the MZIs. In particular a MZI is composed of two 50:50 beam splitters, implemented on-chip using MMI devices, separated by two optical waveguides, each one having a PS. In the ideal case, the 50:50 MMI and the PS matrix representations are:

\begin{equation}
	U_{\text{MMI}}= \frac{1}{\sqrt{2}}\begin{pmatrix}
    1 &i        \\
    i  &1       
    \end{pmatrix}, \quad
	U_{\text{PS}(\zeta_1,\zeta_2)}=\begin{pmatrix}
    e^{2i\zeta_1} &0        \\
    0  & e^{2i\zeta_2}       
    \end{pmatrix}. 
\end{equation}
Consequently, according to standard transfer matrix formalism, the matrix representation of the MZI is:
\begin{equation}\label{eq:matrixMZI}
	U_{\text{MZI}}(\zeta_1,\zeta_2)=ie^{i\left(\zeta_1+\zeta_2\right)}\begin{pmatrix}
    \sin(\zeta_1-\zeta_2) & \cos(\zeta_1-\zeta_2)        \\
    \cos(\zeta_1-\zeta_2)  & -\sin(\zeta_1-\zeta_2)       
    \end{pmatrix}=ie^{iZ}\begin{pmatrix}
    \sin(\zeta) & \cos(\zeta)        \\
    \cos(\zeta)  & -\sin(\zeta)       
    \end{pmatrix}.
\end{equation}
Essentially, each MZI implements a rotation of an angle $\zeta=\zeta_1-\zeta_2$, with a global phase shift of $Z+\pi/2$, with $Z=\zeta_1+\zeta_2$.
Considering the four MZIs represented in Fig.~\ref{fig:chipscheme}, $\zeta_1=\phi_1,\zeta_2=\phi_2$ for the two MZIs that rotate the relative position qubit (first green box from the left in Fig.~\ref{fig:chipscheme}), while $\zeta_1=\theta_1,\zeta_2=\theta_2$ for the two MZIs that rotate the absolute position qubit (second green box from the left in Fig.~\ref{fig:chipscheme}). By implementing two rotations of angles $\phi$ and $\theta$, respectively for the relative and absolute position qubits, the theoretical form of the correlation coefficient $\mathbb{E}(\phi,\theta)$ becomes:
\begin{equation}
   \mathbb{E}(\phi,\theta)=\cos (2 (\phi-\theta))
\end{equation}
and the correlation function $\chi(\phi,\phi',\theta,\theta')$ results to be:
\begin{equation}
	\chi(\phi,\phi',\theta,\theta')
	=\cos (2 (\phi-\theta))
	-\cos (2 (\phi-\theta'))
	+\cos (2 (\phi'-\theta))
	+\cos (2 (\phi'-\theta')).
\end{equation}
By introducing the parameter $\alpha$, such that $2(\phi-\theta)=2(\phi^\prime- \theta^\prime)=-2( \phi - \theta^\prime)=\alpha$, the correlation function can be rewritten as:
\begin{equation}
	\chi(\alpha)=3\cos(\alpha)-\cos (3\alpha).
\end{equation}



\section{Non-idealities in the experimental estimation of $\chi(\phi,\phi',\theta,\theta')$ }\label{section:non-ideal}
Our experimental implementation is affected by a few non-idealities, that in principle could result in a wrong estimation of $\chi(\phi,\phi',\theta,\theta')$.
A first aspect to be considered is the broadband spectrum of emission of the LED. Indeed, the PIC has been designed for $\lambda=730$ nm, so that its performances are optimized at that wavelength. However, the used LED source has a broadband spectrum ($\lambda = 730\pm10$ nm), so that the wavelength dependent behaviour of the integrated optical elements has to be taken into account. For example, the matrix representation of the MMIs becomes:
\begin{equation}
U_{\text{MMI}}(\lambda)=
\begin{pmatrix}
 t(\lambda)& i r(\lambda) \\
 i r(\lambda)  & t(\lambda)\\
\end{pmatrix},
\end{equation}
and consequently the matrix representation of the MZIs is correctly described by:
\begin{equation}
U_{\text{MZI}}(\lambda,\zeta_1,\zeta_2)=
\begin{pmatrix}
 t(\lambda)^2 e^{2 i \zeta_1(\lambda)}-r(\lambda)^2 e^{2 i \zeta_2(\lambda)} & i r(\lambda) t(\lambda) \left(e^{2 i \zeta_1(\lambda)}+e^{2 i \zeta_2(\lambda)}\right) \\
 i r(\lambda) t(\lambda) \left(e^{2 i \zeta_1(\lambda)}+e^{2 i \zeta_2(\lambda)}\right) & t(\lambda)^2 e^{2 i \zeta_2(\lambda)}-r(\lambda)^2 e^{2 i \zeta_1(\lambda)} \\
\end{pmatrix}.
\end{equation}
A second aspect concerns losses. Waveguide propagation losses have to be considered. Moreover MMIs can have insertion losses, meaning that $t^2(\lambda)+r^2(\lambda)\leq 1$. Finally, a third non negligible aspect is represented by current instabilities of the electronics controlling the PSs of the PIC. Indeed, a key feature that must be ensured when a Bell test is performed is that each qubit must be rotated independently by the other\cite{scarani2019}. This is ensured by imposing the same rotation angle on the two MZIs rotating each qubit. More formally, the observables that are considered in a Bell test must be written in product form $U(\phi,\theta)=B(\phi)\otimes A(\theta)$, and current instabilities cannot ensure such necessary product form to be implemented.
Please note that even thermal cross-talk between different PSs of the different MZIs could spoil the wanted product form $U(\phi,\theta)=B(\phi)\otimes A(\theta)$. However, this phenomenon has not be observed, so it will be neglected in the following discussion (see the Section~\ref{section:methods} for further details).
\subsection{Broadband light source}
Here, we consider the wavelength dependence of the transmission and reflection coefficients of the MMIs ($t,r$), as well as of the rotation angles $\zeta_{1(2)}$ set via the MZIs.
Without any loss of generality, we can represent the state $\rho$ of the incoming photons as a convex  superposition of  the following form
   \begin{equation}\rho= \int \rho _\omega d\mu (\omega),\label{state-general}\end{equation}
   for a suitable probability measure $\mu$ over the set of possible frequencies $\omega$, while $\rho _\omega$ represents the state of a monochromatic photon. 
   Since the different elements of the chip have a response which depend explicitly on the frequency $\omega$ of the incoming photons, the overall action of the rotation stage can be described by a family of unitary operators $U(\phi, \theta)_\omega$, in such a way that each component $\rho_\omega$ appearing in the decomposition \eqref{state-general} evolves under the action of the operator $U(\phi, \theta)_\omega$:
   \begin{equation}
   \rho_\omega\mapsto  U(\phi, \theta)_\omega\rho_\omega U(\phi, \theta)^\dag_\omega
   \end{equation}
   and by linearity, the transformation of the general state \eqref{state-general} is given by
   \begin{equation}
   \rho\mapsto  \int U(\phi, \theta)_\omega\rho_\omega U(\phi, \theta)^\dag _\omega d \mu (\omega)\, .
   \end{equation}
   The corresponding quantum probabilities are a convex superposition of the following form
  \begin{equation}
  \mathbb{P}(a,b|\phi,\theta)=\int \mathbb{P}(a,b|\phi,\theta)_\omega d\mu (\omega)
  \end{equation}
  where 
  \begin{equation}
  \mathbb{P}(a,b|\phi,\theta)_\omega=\operatorname{Tr}\left[U(\phi, \theta)_\omega\rho_\omega U(\phi, \theta)^\dag_\omega P_{a}\otimes P_{b}\right].
  \end{equation}
  This is equivalent to say that each monocromatic component $\rho_\omega$ of the generic state \eqref{state-general} is subject to the measurement of a different pair of observables, $A(\phi)_\omega\otimes B(\theta)_\omega$, associated to specific PVMs $\{P^{\theta,\omega}_a\otimes P^{\phi,\omega}_b\}_{a,b}$:
  \begin{equation}
      \mathbb{P}(a,b|\phi,\theta)_\omega=\operatorname{Tr}\left[\rho_\omega P^{\theta,\omega}_a\otimes P^{\phi,\omega}_b\right]=\operatorname{Tr}\left[U(\phi, \theta)_\omega\rho_\omega U(\phi, \theta)^\dag_\omega P_{a}\otimes P_{b}\right].
  \end{equation}
   Analogously, the Bell parameter $\chi$ is given by the convex superposition
   \begin{equation}
       \chi=\int \chi_\omega d\mu(\omega),
   \end{equation}
   with 
   \begin{equation}
       \chi_\omega=\mathbb{E}_\omega(\phi,\theta)-\mathbb{E}_\omega(\phi,\theta')+\mathbb{E}_\omega(\phi',\theta)+\mathbb{E}_\omega(\phi',\theta')
   \end{equation}
where
\begin{equation}
    \mathbb{E}_\omega(\phi),\theta)=\mathbb{P}(a=b|\phi,\theta)_\omega-\mathbb{P}(a\neq b|\phi,\theta)_\omega.
\end{equation}
This means that using a broadband light source results in a correlation function which is a spectrally weighted average. It is important to point out that this fact does not affect the entropy certification protocol (see Section~\ref{section:protocol} and Appendix~\ref{appendix-theoreticalpreliminaries}).

Moreover, we want to stress that here the fact that our light source is non-monochromatic does not require the use of a narrow band optical filter for the actual integrated setup as it is the case for the bulk one reported in \cite{Pasini20}. Indeed, thanks to the fact that the phase delay of MZIs is varied by slightly changing the refractive index of the waveguides ends up in all the rotation operations naturally occurring within the coherence time.

\subsection{Losses}
We can consider two types of losses affecting the PIC:
\begin{itemize}
    \item Waveguide Propagation losses, that can be viewed as a common factor $e^{-\alpha l}$, where $\alpha$ is the attenuation coefficient and $l$ is the length of the path (waveguide) taken by photons;
    \item Optical losses specific of each device, due to its insertion in the PIC.
\end{itemize}
Thanks to the homogeneity of the discrete components and to the  symmetry of the chip, in particular to the nominally equal lengths of the paths corresponding to the four different outputs (detection channels), the overall impact of losses can be modelled by a diagonal operator of the form $L=\gamma I \otimes I$, with $\gamma \in (0,1]$, hence commuting with any operator acting on the 2-qubits Hilbert space.
Therefore, the probability of photon detection per each channel is given by:
\begin{equation}\label{cod-prob-losses}
    \mathbb{P}(a,b|\phi,\theta)=\frac{\operatorname{Tr}\left[U(\phi, \theta)L\rho L^\dag U(\phi, \theta)^\dag P_{a}\otimes P_{b}\right]}{\operatorname{Tr}\left[U(\phi, \theta)L\rho L^\dag U(\phi, \theta)^\dag\right]}\, .
\end{equation}
which can be actually cast in the equivalent form:
\begin{equation}\label{cod-prob-losses_2}
\mathbb{P}(a,b|\phi,\theta)=\operatorname{Tr}\left[U(\phi, \theta)\rho'  U(\phi, \theta)^\dag P_{x}\otimes P_{y}\right],    
\end{equation}

with a different density operator $\rho':=\frac{L\rho L^\dag}{\operatorname{Tr}\left[L\rho L^\dag\right]}.$
In addition, it is even possible to consider the case in which the loss effect depends explicitly on the photons' frequency $\omega$, according to the discussion of the previous subsection.
However, as Eq.~\ref{cod-prob-losses_2} is of the same form of Eq.~\ref{eq:probrot}, we can conclude that this issue does not directly affect the estimate of the correlation function $\chi$.

\subsection{Current instabilities}
In all cases in which electrical currents applied to PSs do not precisely correspond to the desired nominal values, an error $(\delta\zeta)$ is introduced for each PS. As a general consequence, the matrices representing the rotation operators implemented by the MZIs are no longer in product form, but they have to be written as follows:
\begin{equation}
	\begin{aligned}
	&U_{\text{real}}(\phi_1,\phi_2,\delta\phi_1,\delta\phi_2,\delta\phi_3,\delta\phi_4)=
	P1 \otimes U_{\text{MZI}}(\phi_1+\delta\phi_1,\phi_2+\delta\phi_2)+
	P2 \otimes U_{\text{MZI}}(\phi_1+\delta\phi_3,\phi_2+\delta\phi_4);\\
	&U_{\text{real}}(\theta_1,\theta_2,\delta\theta_1,\delta\theta_2,\delta\theta_3,\delta\theta_4)=
	U_{\text{MZI}}(\theta_1+\delta\theta_1,\theta_2+\delta\theta_2) \otimes P1 +
	 U_{\text{MZI}}(\theta_1+\delta\theta_3,\theta_2+\delta\theta_4) \otimes P2;\\
	&P_1=
	\begin{pmatrix}
		1 & 0 \\
		0 & 0
	\end{pmatrix}, \quad
	P_2=
	\begin{pmatrix}
		0 & 0 \\
		0 & 1
	\end{pmatrix}.
	\end{aligned}
\end{equation}
The terms $\delta\phi_1,\delta\phi_2,\delta\phi_3,\delta\phi_4,\delta\theta_1,\delta\theta_2,\delta\theta_3,\delta\theta_4$ are all different and appear both in the rotation angles $\phi=\phi_1-\phi_2$ and $\theta=\theta_1-\theta_2$, as well as in the global phase terms imposed by the MZIs $\Phi=\phi_1+\phi_2$ and $\Theta=\theta_1+\theta_2$.
In this situation the matrix $U_{\text{PS}}$ is influenced by the experimental error as:
\begin{equation}
	U^{\text{real}}_{\text{PS}(\zeta_1,\zeta_2)}=\begin{pmatrix}
    e^{2i(\zeta_1+\delta\zeta_1)} &0        \\
    0  & e^{2i(\zeta_2+\delta\zeta_2)}       
    \end{pmatrix}. 
\end{equation}
In general, due to these current instabilities, the effective angle of rotation is different for each PS of each MZI. 
For clarity, we focus our discussion on the rotation angle $\phi$, for which $\delta\zeta_1=\delta\phi_1,\delta\zeta_2=\delta\phi_2,\delta\zeta_3=\delta\phi_3,\delta\zeta_4=\delta\phi_3$. Then, the same arguments can be applied to current-related non-idealities affecting the other rotation angle $\theta$.
We start by representing the real operator describing the rotation of the related qubit in the following form:
\begin{multline}
    U^{\text{real}}(\phi_1,\phi_2,\delta\phi_1,\delta\phi_2,\delta\phi_3,\delta\phi_4 )= \\ P_1\otimes U_{\text{MMI}}U^{\text{real}}_{\text{PS}(\phi_1,\phi_2,\delta \phi_1, \delta\phi_2)}U_{\text{MMI}}\\ +P_2\otimes U_{\text{MMI}}U^{\text{real}}_{\text{PS}(\phi_1,\phi_2,\delta \phi_3, \delta\phi_4)}U_{\text{MMI}}
\end{multline}
This problem is addressed by using the same techniques reported in \cite{Mazzucchi2021,Leone2022}: we look for an ideal operator that can be written as $I\otimes U^{\text{ideal}}$ whose distance from $U^{\text{real}}$ is the smallest possible, i.e. an operator minimizing the distance from $U^{\text{real}}(\phi_1,\phi_2,\delta\phi_1,\delta\phi_2,\delta\phi_3,\delta\phi_4 )$ according to the Hilbert-Schmidt norm.
Without loss of generality, we can represent $U^{\text{ideal}}$ as the product $U^{\text{ideal}}=U_{\text{MMI}}W^{\text{ideal}}U_{\text{MMI}}$, for a suitable unitary operator $W^{\text{ideal}}$. 
In the general case, the real operator describing the rotation of one qubit by an angle $\phi$ can be represented as 
\begin{multline}
    U^{\text{real}}(\phi_1,\phi_2,\delta\phi_1,\delta\phi_2,\delta\phi_3,\delta\phi_4 )\\ =(I\otimes U_{\text{MMI}})( I\otimes U^{\text{ideal}}_{\text{PS}}(\phi_1,\phi_2))D(\delta \phi_1, \delta\phi_2,\delta\phi_3,\delta\phi_4)(I\otimes U_{\text{MMI}}),
\end{multline}
where
\begin{equation}
    D(\delta \phi_1, \delta\phi_2,\delta\phi_3,\delta\phi_4) =
\begin{pmatrix}
e^{2i\delta \phi_1}&0 &0 &0\\
0 &e^{2i\delta \phi_2} &0&0\\
0 & 0 & e^{2i\delta \phi_3}&0\\
0 &0 &0 &e^{2i\delta \phi_4}
\end{pmatrix}.
\end{equation}
Similarly, we can rewrite the factorized unitary operator $U^{\text{ideal}}$, which minimizes the Hilbert-Schmidt distance from $U^{\text{real}}$, as 
\begin{equation}
    U^{\text{ideal}}=I\otimes \left(U_{\text{MMI}}U_{\text{PS}}(\phi_1,\phi_2)V^{\text{ideal}}U_{\text{MMI}}\right)
\end{equation}
for a suitable unitary operator $V^{\text{ideal}}$, where $W^{\text{ideal}}=U_{\text{PS}}(\phi_1,\phi_2)V^{\text{ideal}}$. By exploiting the representation of a generic element of SU(2) as the exponential of a Pauli vector, the generic unitary operator $V^{\text{ideal}}$ will be written in the form 
\begin{equation}
    V^{\text{ideal}}=e^{i\varphi}e^{i\vartheta \hat n \cdot \sigma}=e^{i\varphi}\left(\cos\vartheta I +i\sin \vartheta \hat n \cdot \sigma\right),
\end{equation}
for some $\varphi, \vartheta \in [0,2\pi)$, and $\hat n\in \mathbb{R}^3$, $\|\hat n\|=1$.
It can be proven (see Appendix~\ref{appendix-2}) that for 
\begin{equation}
\hat n=(0,0,1), 
\quad
\varphi=\frac{\delta\phi_1+\delta\phi_3+\delta\phi_2+\delta\phi_4}{2}, \quad 
\vartheta= \frac{\delta\phi_1+\delta\phi_3-\delta\phi_2-\delta\phi_4}{2}
\end{equation}
the Hilbert-Schmidt distance from $U^{\text{real}}$ is minimum.
We remark that the same reasoning can be applied also for current instabilities affecting the angle $\theta$.

Consequently, it is necessary to evaluate 
\begin{equation}\label{eq:echi}
e_\chi=
\max_{\{\phi,\phi',\theta,\theta',\rho\}}\left| \chi^{\text{ideal}}(\phi,\phi',\theta,\theta')-\chi^{\text{real}}(\phi,\phi',\theta,\theta') \right|,
\end{equation}
which represents a correction term that has to be applied to the experimental correlation function $\chi^{\text{real}}$. It takes into account the fact that with $U^{\text{real}}$ we are not exactly applying a factorized operator to the state $\rho$ and so we are making the error $e_\chi$. To estimate such a correction term we use the numerical approach detailed in \cite{Leone2022}.
Note that in the construction of $\chi(\phi,\phi',\theta,\theta')$ four correlation coefficients $\{\mathbb{E}_i\}_{i=1..4}$ are considered, each of them having different values of the errors $\{\{\delta\phi_{i,j},\delta\theta_{i,k}\}_{j,k=1..4}\}_{i=1..4}$.To simplify the evaluation, we evaluate $\{e_{\chi,\xi}\}_{\xi=1..4}$ considering that the four $\{\mathbb{E}_i\}_{i=1..4}$ in $\chi$ have the same errors $\{\delta\phi_{\xi,j},\delta\theta_{\xi,k}\}_{j,k=1..4}$ of the correlation coefficient $\mathbb{E}_{xi}$.

\subsection{Other sources of non idealities}
Other non-idealities in our experimental setup come from the detectors (dead-time, afterpulsing and dark counts), especially because no randomization of the input sequence of measurement operations necessary to estimate the probabilities $\{\mathbb{P}(a,b|\phi,\theta)\}_{\phi,\theta,a,b}$ is performed. In this particular implementation we neglect such non-idealities because in \cite{Leone2022} it has been observed that, for a photon flux of $\simeq 200$ kHz, the corrections to the probabilities are negligible. As a lower flux of photons $\simeq 120$ kHz is here used (mainly because of LED-to-fiber low coupling efficiency), we can safely neglect such a correction factor.
\section{Methods}\label{section:methods}
The PIC has been fabricated in the cleanroom of Fondazione Bruno Kessler.
The waveguide core ($300$ nm-thick and $700$ nm-wide) is made of Silicon Oxynitride ($\text{SiO}_x\text{N}_y$) and the bottom and top claddings are made, respectively, of thermally grown Silicon Oxide ($\text{Si}\text{O}_2$) and Borophosphosilicate glass (BPSG). The design wavelength of the integrated components is $730$ nm and mode polarization is transverse electric (TE). The linear characterization of the building-blocks of the PIC, namely MMIs and CRs, has been carried out using a spectrally filtered supercontinuum laser emitting from $300$ nm to $2000$ nm, and details about their performances can be found in Appendix \ref{appendix-1}. 

Our experiment is performed using an attenuated commercial LED at $730\pm10$ nm. The light coming from the LED is collimated using an objective and polarized using a Glan-Thompson polarizer. A fiber-port is used to couple light inside an optical fiber, where light is attenuated by means of a variable optical attenuator and polarization is controlled and set to be TE at the output. The input-output coupling is performed using a standard fiber array. A power supply is used to control the different phase shifters present on the PIC. 

Prior to our measurements, a mandatory calibration of integrated MZIs has been performed using a Ti:Sapphire laser tuned at $730$ nm. In particular, the phase-power relation $\phi=\phi(W)$ of the different MZIs is retrieved by fitting operations of the transmitted optical power using the following equations:
\begin{equation}
    I_{\text{out}1}=a\cos(b W+ d)^2 + c
\end{equation}
\begin{equation}
    I_{\text{out}2}=a\sin(b W+ d)^2 + c
\end{equation}
corresponding to the two output ports of an MZI, where $a,b,c,d$ are fitting parameters. This allows to reconstruct the relation $\phi=\phi(W)$ for each MZI by means of the linear function $\phi(W)=bW+d$.
Please note that, to eliminate the thermal crosstalk between different MZIs, each PS is encapsulated between trenches (areas in which the core and cladding materials are removed) to limit the heat propagation inside the chip. This and the low thermal conductivity of the Silicon Oxide, strongly limit the thermal crosstalk, making it negligible during the experiment.

The phase $\xi=\xi_1-\xi_2$ of the generation stage is obtained by another fitting operation, applying the correct calibration to each MZI. Here an issue is due to the absence of additional compensation phase shifters after the MZIs rotating the relative-position qubit. In this situation, the value of $\xi$ to fix the phase difference between the $|N\rangle$ components of the state basis is different with respect to the one for the $|F\rangle$ components. For this reason the following strategy is introduced. First, we set the value of $\xi$ necessary to have the expected phase relation between the $|F\rangle$ components, and only counts from channels $|UF\rangle$ and $|DF\rangle$ are acquired. Second, the experiment is repeated acquiring only counts from channels $|UN\rangle$ and $|DN\rangle$, by changing $\xi$ to the value that adjusts the phase relation of the $|N\rangle$ components too.

Finally, the measurements to demonstrate our certified QRNG integrated device are performed using four SPADs (Excelitas), whose efficiencies have been equalized using fiber-coupled variable optical attenuators, and whose electrical outputs are sent to a time-tagging electronics (Swabian Instruments) connected to a PC in order to count single-photon detection events for each channel over time. The following procedure has been followed. First, an angle of rotation $\theta$ of the absolute-position qubit is fixed and a sweep over $\phi$ angles of relative-position is performed. For each couple $(\theta,\phi)$ of angles, a 1 s time window is acquired with a time-bin set at 1 $\mu$s for the time tagger. Then, the angle $\theta$ is changed and a sweep of the other angle $\phi$ is again performed. Time bins with no detection events are discarded, while whenever multiple photon detection events are registered at different SPADs within the same time bin, one single event is randomly assigned to one of the four two-qubit states by means of a pseudo random numbers generator.

\section{Experimental demonstration of the presence of entanglement}\label{sec:expdemonstration}
\begin{figure}[h!]
 \centering
	\includegraphics{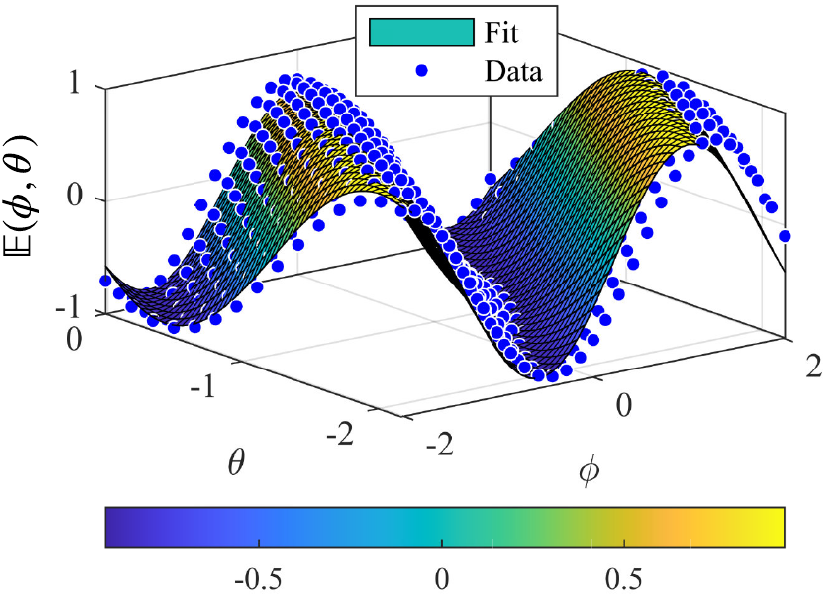}
	\caption{Experimental correlation coefficients $\mathbb{E}\left(\phi,\theta\right)$) (blue dots) with the related fit (colored surface), according to Eq. \ref{eq:fitcor}. $\phi$ is the rotation angle of the relative-position qubit, while $\theta$ is the rotation angle of the absolute-position qubit. Color bar refers to the value of $\mathbb{E}$.}
	\label{fig:chipcorrelationcoeffnoncorr}
\end{figure}

The experimental demonstration of the presence of entanglement is performed by fixing the angle $\theta \in [-2,0] \text{rad}$ and performing a sweep of the rotation angle $\phi \in [-2,2] \text{rad}$. An acquisition 1 s long with a time bin of $1\ \mu$s is performed for each $(\phi,\theta)$. The average photons flux is $\simeq 120$ kHz. In this way, we estimate the conditional probabilities $\mathbb{P}(a,b|\phi,\theta)$ as the empirical frequencies of the counts:
 \begin{equation}
     \mathbb{P}(|ab\rangle|\phi,\theta)=\frac{N_{|ab\rangle}}{\sum N_{|ab\rangle}},
 \end{equation}
 where $N_{|ab\rangle}$ is the number of photons detected in the state $|ab\rangle$ and $\sum N_{|ab\rangle}$ is the total number of photons detected.
Note that no-detection events are eliminated from the detection sequence, while multiple detection events are randomly assigned to only one of the detection channels using a pseudo random number generator. The experimental correlation coefficients $\mathbb{E}\left(\phi,\theta\right)$ are reported in Fig. \ref{fig:chipcorrelationcoeffnoncorr} as data points. The colored surface reported in Fig.~\ref{fig:chipcorrelationcoeffnoncorr} represents the theoretical correlation coefficient $\mathbb{E}\left(\phi,\theta\right)$ in the case of non ideal 50:50 beam splitters with transmission and reflection power coefficients of $T=40\%$ and $R=60\%$, respectively (see Appendix~\ref{appendix-1} for the results of our experimental characterization). This results:
 \begin{equation}\label{eq:fitcor}
 \begin{aligned}
		\mathbb{E}\left(\phi,\theta\right)=&\eta(5+(-48)\sqrt{6}+(-24)(5+2\sqrt{6})\cos(2\phi)+\\
		&+(-24)(5+2\sqrt{6})\cos(2\theta)+\\
		&+48(30+(-13)\sqrt{6})\cos(2(\phi+\theta))+\\
		&+288(5+2\sqrt{6})\cos(2(\phi-\theta))),
\end{aligned}
\end{equation}
where $\eta=\frac{1}{3125}\simeq 3.2 \times 10^{-4}$. Instead, by a fit of the experimental data, we obtain $\eta= (3.01\pm0.04) \times 10^{-4}$. The difference between the theoretical and the fitted $\eta$ valus could be explained by the broadband spectrum of the LED source, since for $\lambda\neq730$ nm the parameters of the MMI are slightly different.
\begin{figure}[h!]
\centering
	\includegraphics{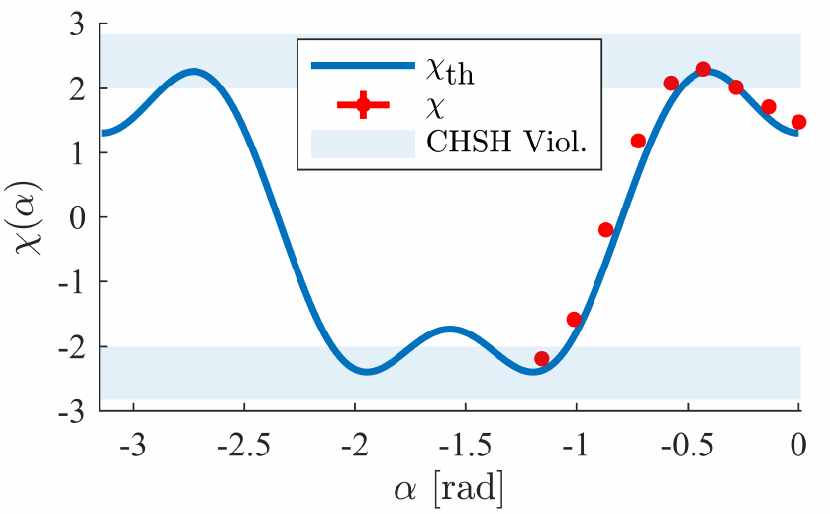}
	\caption{Experimental demonstration of the violation of the Bell inequality. Data points (red dots) with their error bars (smaller than the size of the data points) and the theoretical curve (blue line) of the $\chi$ correlation function, both with respect to the parameter $\alpha$. In cyan, the areas corresponding to violation of Bell inequality. Due to a failure of the wire bonding of one PS of one MZI, it was possible to acquire only data points in a limited range of $\alpha$.}
    \label{fig:chifun}
\end{figure}
Using these correlation coefficients, it is possible to construct the correlation function $\chi(\phi,\phi',\theta,\theta')$. By making the choice $\phi=-\alpha,\phi'=\alpha,\theta=0, \theta'=2\alpha$,
the correlation function can be plotted as a function of the parameter $\alpha$. The resulting $\chi(\alpha)$ function is reported in Fig.~\ref{fig:chifun} as a blue curve (obtained starting from Eq.~\ref{eq:fitcor} with the fitted value of $\eta$), while measured data are plot as red dots. A good agreement can be observed. It can also be noticed that the theoretical form of the correlation function does not reach the maximum attainable value of $2\sqrt{2}$: this is caused by the non-ideal transmission and reflection coefficients of the MMIs. However, a clear violation of the CHSH inequality can be observed, reaching a maximum value $\chi^+=2.297\pm0.004$ and as minimum $\chi^-=-2.181\pm0.004$, meaning that single photon entangled states have been generated. Moreover, it is important to underline that the choice $\phi=-\alpha,\phi'=\alpha,\theta=0, \theta'=2\alpha$ gives the maximum achievable Bell inequality violation only if the considered entangled state is exactly $|\phi^+\rangle$. However, due to the fact that the MMI-based beam splitters do not have a 50:50 branching ratio, in our experimental implementation the generated state is actually different. For this reason, it is a better strategy to consider every possible combination of $(\phi,\phi',\theta,\theta')$ between the measured $\mathbb{E}(\phi,\theta)$  to look for a better violation of the inequality.
Using this approach, we obtain respectively $\chi^+= 2.697\pm0.004$ for $(\phi_0=-0.576 \pm 0.002,\phi_1=-1.445 \pm 0.002,\theta_0=-1.11\pm 0.02,\theta_1=-1.87\pm 0.02)$ and $\chi^-=-2.668\pm0.004$ for $(\phi_0=-1.589 \pm 0.002,\phi_1=0.863 \pm 0.002,\theta_0=-0.35\pm 0.02,\theta_1=-1.27\pm 0.02)$. Error bars on $\chi$ data points are obtained by dividing each $1$ s time sequence in intervals of $0.2$ s. For each interval, the correlation function is evaluated and the final uncertainty is obtained as the standard error related to the number of intervals. Lastly, it is necessary to correct the values $\chi^{\pm}$ with the terms $e_{\chi^{\pm}}$ (Eq.~\ref{eq:echi}). Using the values of current instabilities-related errors reported in Table~\ref{tab:errorsphase}, we obtain $e_{\chi^+}=0.092$ for $\chi^+$ and $e_{\chi^-}=0.077$ for $\chi^-$.
As a result, the  corrected values of the correlation functions are $\chi^+= 2.605\pm0.004$ and $\chi^-=-2.591\pm0.004$.
\begin{table}[h!]
    \centering
    \begin{tabular}{c||c|c|c|c|c|c|c|c}
         &$\delta\phi_1$  & $\delta\phi_2$ & $\delta\phi_3$ & $\delta\phi_4$ & $\delta\theta_1$ & $\delta\theta_2$ & $\delta\theta_3$ & $\delta\theta_4$ \\
        \hline
        \hline
        $\chi^+$ & 0.000 & 0.011 & -0.004 & -0.006 & 0.068 &  0.216 & 0.036 & 0.215\\
        $\chi^-$ & 0.002 & 0.004 & 0.007 & -0.006 & 0.068 &  0.187 & 0.036 & 0.180
    \end{tabular}
    \caption{Errors on the rotation angles $\phi$ and $\theta$ in the estimation of $\chi^{\pm}$. The standard error for each values is $\delta=0.003$ and it is obtained through repeated measurements. Note that the errors on $\theta$ are greater that the errors on $\phi$. This is due to the fact that one of the heaters enabling the $\theta$ rotation was not working.}
    \label{tab:errorsphase}
\end{table}

\section{From entanglement to quantum-certified random numbers}\label{section:protocol}
Recently, a few of us demonstrated that it is possible to obtain a semi-device independent (SDI) randomness certification scheme starting from SPE states of momentum and polarization using a bulky experimental implementation\cite{Mazzucchi2021,Leone2022}. 
Therefore, here we use the PIC reported in Fig.~\ref{fig:chipscheme} as an SDI-QRNG, certified by the evaluation of the correlation function $\chi$. A random number is produced every time that, fixing a couple of angle $(\phi_i,\theta_j)$, we detect a single photon on one of the four SPADs at the output of the chip. The state in which the rotated wavefunction has collapsed represents the random digit.
The generated raw random sequence is then composed by the four time sequences of  detection events corresponding to the angles $\{(\phi_i,\theta_j)\}_{i,j=0,1}$ used to evaluate $\chi$.
We use an entropy certification protocol similar to the one reported in \cite{Mazzucchi2021,Leone2022}.
First of all, we recall the hypotheses over which our certification scheme is based:
\begin{enumerate}
    \item SPADs are characterized;
    \item a characterization of all the MZIs present on the chip is available;
    \item the power supply that drives the currents to the PSs is error-prone and it is not controlled by an adversary;
    \item the generation and measurement parameters must be stable during the acquisition time.
\end{enumerate}
Hypothesis 1 is necessary to reconstruct the probabilities $\mathbb{P}(a,b|\phi,\theta)$. Indeed, as soon as the couple of angles $(\phi,\theta)$ is not randomized for every round of the experiment, an adversary could induce detector clicks in a deterministic way for mimicking the violation of a Bell inequality.
Hypotheses 2-3, instead, are necessary to be able to set the same angle $\phi$ and $\theta$ on the different couples of MZIs. In particular, hypothesis 2 is required to take into account the fact that the starting phase of each MZI is not exactly 0, while hypothesis 3 is necessary to set the correct currents at phase shifters. Finally, hypothesis 4 is necessary to rule out any possible measurement basis-dependent change of the input state\cite{Leone2022}.
Our certification protocol is independent of the particular form of the input state $\rho$. In particular, for the four sequences of  detection events corresponding to the angles $\{(\phi_i,\theta_j)\}_{i,j=0,1}$, the related conditional guessing probability can be lower bounded by using\cite{Mazzucchi2021,Leone2022}:
\begin{equation}\label{eq:corrected_H}
	\mathbb{P}_{\text{guess}}(a,b|\phi_i,\theta_j) \leq \frac{1}{2}+\frac{1}{2}\sqrt{2-(|\chi_{\text{real}}|-e_\chi)^2/4} + e_P, 
\end{equation}
where $\chi_{\text{real}}$ is the correlation function in the CHSH form estimated from the experimental data. $e_\chi$ represents the correction term previously introduced in Section~\ref{section:non-ideal}. We stress that the maximization procedure for $e_\chi$ is run over every possible combination of angles $\{\phi,\phi',\theta,\theta'\}$, in such a way to map every possible operator $U^{\text{real}}$, and over every possible state $\rho$.
$e_P$ is another numerical correction term, which has the same meaning of $e_\chi$ but for probabilities instead: it represents an upper bound for the difference between the ideal probabilities obtained by measuring factorized observables and the estimated probabilities obtained in the presence of the non-idealities here considered.
$e_P$ is estimated in the same way as $e_\chi$ by using the numerical methods described in \cite{Leone2022}.
We remind the reader that, with respect to the result reported in \cite{Leone2022}, here we are neglecting the Markovian correction to the guessing probability $\mathbb{P}_{\text{guess}}$, introduced to take into account memory effects due to detectors non-idealities, such as afterpulsing and dead time. Indeed, as this work is interested by a lower flux compared to the one reported in that work, the effect of that correction is actually negligible.
Note that Eq. \ref{eq:corrected_H} is valid even considering the broadband spectrum of our LED source (see the demonstration in Appendix \ref{appendix-theoreticalpreliminaries}).

\begin{figure}[h!]
\centering
	\includegraphics{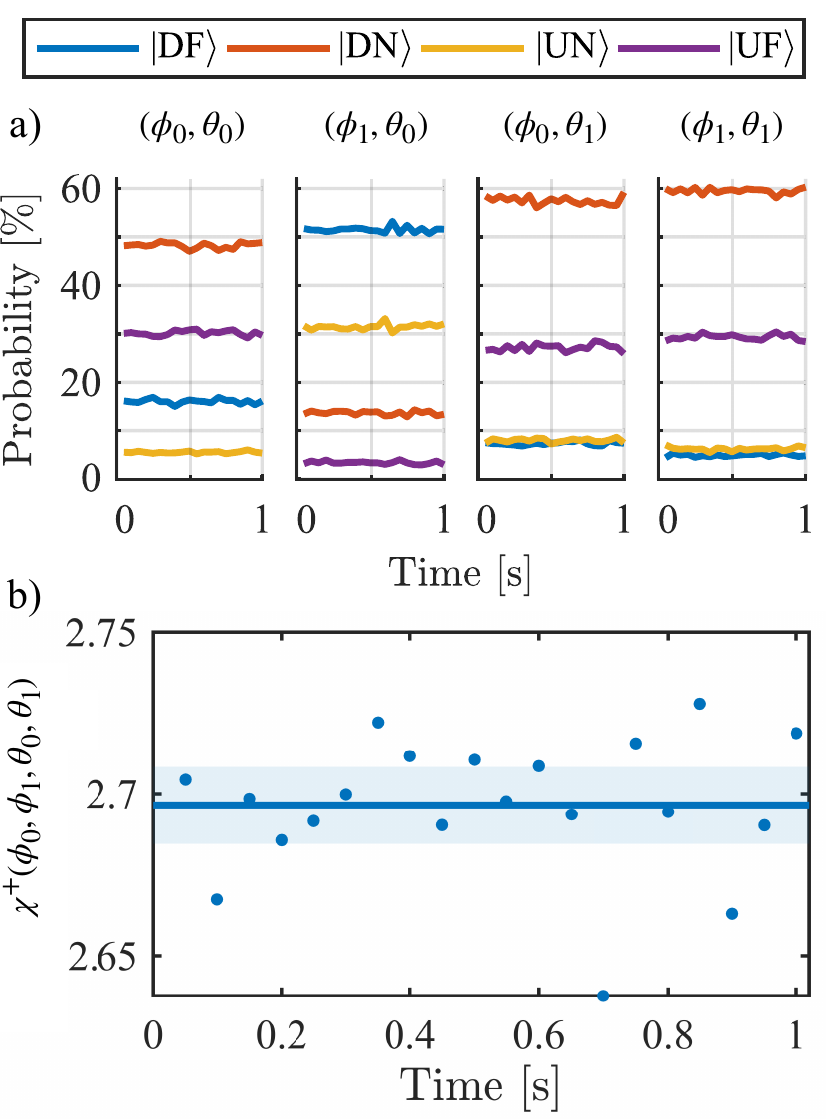}
	\caption{a) Probabilities of each measurement outcome as a function of time (blue $|DF\rangle$, red $|DN\rangle$, yellow $|UN\rangle$ and purple $|UF\rangle$) for the four couples of angles $(\phi_0,\theta_0)$, $(\phi_1,\theta_0)$, $(\phi_0,\theta_1)$, $(\phi_1,\theta_1)$ of $\chi^+$. The estimates have been done considering time intervals of 50 ms.
    b) Dots: corresponding values of $\chi^+$ as a function of time. Solid line: mean value of $\chi^+$. Dashed region: $99\%$ confidence interval.}
    \label{fig:probchiplus}
\end{figure}

\begin{figure}[h!]
\centering
	\includegraphics{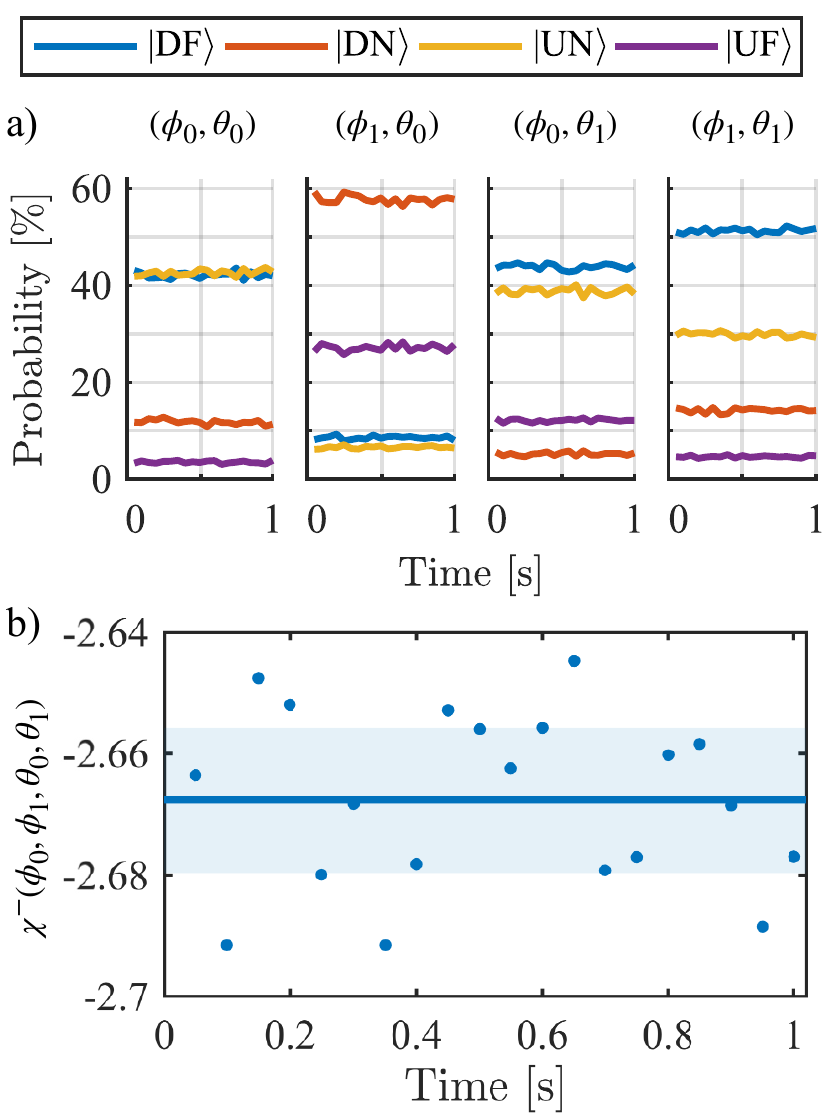}
	\caption{
    a) Probabilities of each measurement outcome as a function of time (blue $|DF\rangle$, red $|DN\rangle$, yellow $|UN\rangle$ and purple $|UF\rangle$) for the four couples of angles $(\phi_0,\theta_0)$, $(\phi_1,\theta_0)$, $(\phi_0,\theta_1)$, $(\phi_1,\theta_1)$ of $\chi^-$. The estimates have been done considering time intervals of 50 ms.
    b) Dots: corresponding values of $\chi^-$ as a function of time. Solid line: mean value of $\chi^-$. Dashed region: $99\%$ confidence interval.}
    \label{fig:probchiminus}
\end{figure}

To experimentally demonstrate the generation of certified random numbers we select the couple of angles giving the maximum and minimum violation of the Bell Inequality, $\chi^+$ and $\chi^-$ respectively, as reported in Section~\ref{sec:expdemonstration}.
The time traces of the estimated probabilities used for computing the correlation functions $\chi^+$ and $\chi^-$ are reported in Fig.~\ref{fig:probchiplus}(a) and Fig.~\ref{fig:probchiminus}(a). These are obtained as 50 ms long averages. As it can be observed, the probabilities are quite stable during the entire acquisition time. The values of $\chi$ corresponding to these time intervals are then reported in Fig.~\ref{fig:probchiplus}(b) for $\chi^+$ and Fig.~\ref{fig:probchiminus}(b) for $\chi^-$, together with their mean value and related $99\%$ confidence interval (blue shaded region).
Considering these values of violation of Bell inequality and the corresponding values of $e_P$ in the two cases, which are respectively $e_P^{\chi^+}=0.02$ and $e_P^{\chi^-}=0.014$, we obtain the following guessing probabilities:
$$
\mathbb{P}_{\text{guess}}(a,b|\phi_x,\theta_y)^+=0.796 \pm 0.002,
$$
$$
\mathbb{P}_{\text{guess}}(a,b|\phi_x,\theta_y)^-=0.798 \pm 0.002.
$$
By applying the formula\cite{Koning09}:
\begin{equation}
    H_{\text{min}}=-\log_2\left[\mathbb{P}_{\text{guess}}(a,b|\phi_x,\theta_y)\right],
\end{equation}
we get
$$
H_{\text{min}}^+=(33.0\pm0.4)\%
$$
$$
H_{\text{min}}^-=(32.6\pm0.4)\%
$$
Having used an average rate of 120 kHz for each acquisition, the final rate of our quantum-certified SDI-QRNG is given by $(120 \times H_{\text{min}})$ kHz which, in the best case (namely $\chi^+$), gives a generation rate of $\simeq 40$ kHz. These rate does not consider the randomness extraction procedure\cite{Nisan99}.

\section{Conclusion}\label{section:conclusion}

Single-photon path entanglement has been demonstrated in a photonic integrated circuit (PIC) by measuring the Bell inequality violation in the CHSH formulation. Photons are generated in an off-chip LED, are manipulated in a PIC fabricated in a silicon foundry and are detected by using off-chip silicon SPADs. The PIC is based on simple and well-known optical components, ie. multi-mode interferometers (MMIs), phase shifters (PSs) based on the thermo-optic effect, Mach-Zehnder interferometers (MZIs) and crossings (CRs). Experimental data are well fitted by a theoretical model which considers the non-idealities of our setup, mainly related to unbalanced beam splitters (MMIs). A large violation of the CHSH inequality is reported ($2.605 \pm 0.004$) by recursively trying each possible combination of the acquired correlation coefficients.

These SPE states are used to demonstrate a certified QRNG working accordingly to a semi-device independent certification protocol similar to the one in \cite{Mazzucchi2021,Leone2022}. Since in \cite{Mazzucchi2021,Leone2022} discrete optical components and different degrees of freedom were used, a few assumptions are here different. Specifically, our present approach is based on the knowledge of the single-photon detectors as well as of the integrated MZIs. In addition, the power supply controlling the heaters in the MZIs must be trusted, even if error-prone, as well as the set phase delays must be stable during the acquisition time. No hypothesis are needed on the input state. 
Under these assumptions, we are able to certify a maximum value of the quantum min-entropy $H_{\text{min}}=(33.0\pm0.4)\%$, which is one order of magnitude larger than that obtained with a bulk setup \cite{Leone2022}. This large improvement is due to the smaller correction terms $e_{\chi}$ and $e_{P}$ of the integrated QRNG because of the smaller non-idealities in the integrated photonic components with respect to the discrete optical components. The measured QRNG generation rate ($\simeq 40$ kHz) is strongly limited by the  actual coupling efficiency of the LED to the optical fiber and to the PIC which decreases the photon flux in the PIC. In future experiments, a better coupling scheme based on e.g. optimized grating couplers or direct LED bonding could significantly improve the coupling efficiency. Then, the other limit will be given by the saturation rate of the SPAD detectors. Considering to work at the linearity limit of SPADs, i.e. 1 MHz rate, the achievable certified random bit rate is of the order of 330 kHz. To further increase such a value, it is necessary to use engineered detectors, e.g. by multiplexing more SPADs in a single detector. The use of Silicon photomultipliers, which are arrays of SPADs, could be a viable solution.

To conclude, a PIC able to generate quantum certified random numbers using single-photon path-entangled states represents a further step to move semi-device independent QRNGs from the lab to real-world applications.

\begin{backmatter}
\bmsection{Funding}
This project has been supported by Q@TN, the joint lab between University of Trento, FBK- Fondazione Bruno Kessler, INFN-National Institute for Nuclear Physics and CNR-National Research Council. This project has received funding from the European Union’s Horizon 2020 research and innovation programme under grant agreements No 820405 Project QRANGE and No 899368 Project EPIQUS. 


\bmsection{Disclosures}
L.P., V.M., S.M. declare the following competing interests: a patent has been filed on single-photon entanglement.

\bmsection{Data availability} Data underlying the results presented in this paper are not publicly available at this time but may be obtained from the authors upon reasonable request.

\appendix
\section{Appendix: Characterization of the integrated optical devices}\label{appendix-1}

The characterization of CRs and MMIs has been performed by means of a supercontinuum laser and two tapered optical fibers for in and out coupling to the PIC. Input laser light is TE-polarized using two half-wave plates and one quarter-wave plate. Spectrally resolved detection is done with an optical spectrum analyzer. The test structure for CR is a sequence of 150 CRs not-equidistant to avoid any Fabry-Perot effect.
Transmission spectrum of a single CR normalized to a reference straight waveguide is shown in Fig.~\ref{fig:Mis_Cr_Det}: a measured power transmission coefficient of $\simeq 98\%$ is reported at $730$ nm. 
\begin{figure}[h!]
\centering
	\includegraphics{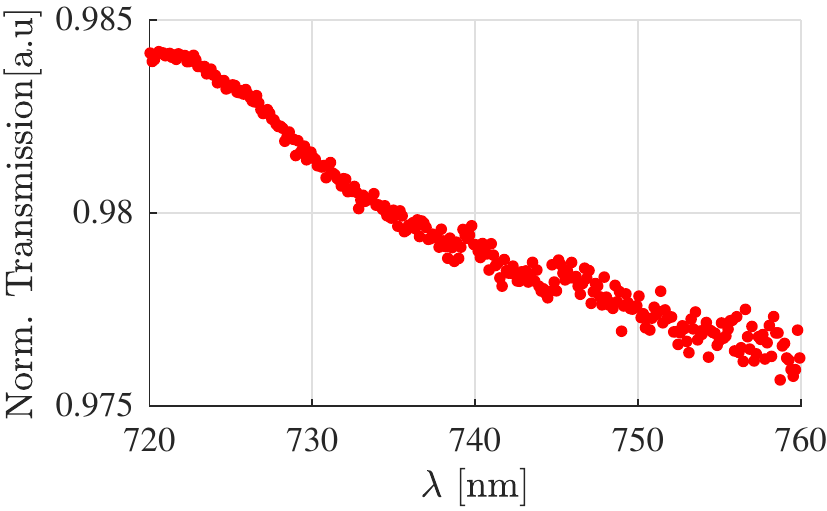}
	\caption{Measured transmission spectrum of a single crossing in SiON.}
	\label{fig:Mis_Cr_Det}
\end{figure}

For MMIs, optical power at the two output ports is collected for light entering at both inputs. The experimental data in Fig.~\ref{fig:Mis_MMI_Det} are again normalized with respect to a straight waveguide transmission spectrum: $Pij$ indicates the normalized power coefficient of output $i$ with respect to input $j$. At $730$ nm, the transmission coefficients $P11$ and $P22$ are measured to be $\simeq 40\%$, while the transmission coefficients $P12$ and $P21$ are $\simeq 60\%$. These are the characterized values entering into hypothesis 2 of our certification protocol.

\begin{figure}[h!]
\centering
 \includegraphics{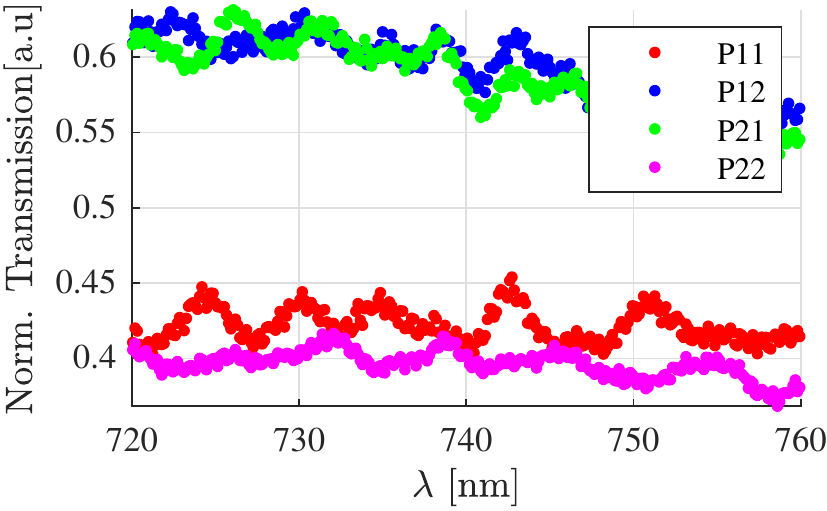}
	\caption{Measured transmission spectra of a MMI-based integrated beam splitter made of SiON.}
	\label{fig:Mis_MMI_Det}
\end{figure}

\section{Appendix: Explicit computation of the values of $\hat{n},\phi,\theta$ that minimize the Hilbert-Schmidt distance}\label{appendix-2}
The square of the Hilbert-Schmidt distance (HS-distance) between $U^{\text{ideal}}$ and $U^{\text{real}}$ is given by:
\begin{align*}
   & \operatorname{Tr} [( U^{\text{real}}-U^{\text{ideal}})( U^{\text{real}}-U^{\text{ideal}})^\dag]\\
   = &8-\operatorname{Tr}[D(\delta \phi_1, \delta\phi_2,\delta\phi_3,\delta\phi_4)(I\otimes V^{\text{ideal}})^\dag+h.c.]\, ,
\end{align*}
hence, one has to find the optimal parameters $\varphi, \vartheta, \hat n$ maximizing the term $\operatorname{Tr}[D(\delta \phi_1, \delta\phi_2,\delta\phi_3,\delta\phi_4)(I\otimes e^{i\varphi}e^{i\vartheta \hat n \cdot \sigma})^\dag+h.c.]$.
By direct computation this term is given by:
$$\operatorname{Tr}[V^{\text{ideal} }U_1^\dag +U_1(V^{\text{ideal} })^\dag]+\operatorname{Tr}[V^{\text{ideal} }U_2^\dag +U_2(V^{\text{ideal} })^\dag]$$
with 
\begin{align*}
    U_1&=\begin{pmatrix}
e^{2i\delta \phi_1} & 0\\
0 & e^{2i\delta \phi_2}
\end{pmatrix},\\
U_2&=\begin{pmatrix}
e^{2i\delta \phi_3} & 0\\
0 & e^{2i\delta\phi_4 }
\end{pmatrix}\, .
\end{align*}   
For a particular phase shifter operator of the form
$$U(\alpha , \beta)=\begin{pmatrix}
e^{2i\alpha} & 0\\
0 & e^{2i\beta }
\end{pmatrix}=e^{i(\alpha +\beta)} e^{i(\alpha -\beta )\sigma_z} $$
and a generic unitary operator $V=e^{i\varphi}e^{i\vartheta \hat n \cdot \sigma}$, by using the composition rule in $SU(2)$ one can easily obtain the following formula 
\begin{equation}
    \operatorname{Tr}[U(\alpha , \beta)V^\dag +VU^\dag(\alpha , \beta)]=4\cos(\varphi-\alpha -\beta)\left( \cos\vartheta \cos(\alpha -\beta)+n_z\sin\vartheta \sin (\alpha-\beta)\right)
\end{equation}
In particular, we have:
\begin{multline}
    \operatorname{Tr}[V^{\text{ideal} }U_1^\dag +U_1(V^{\text{ideal} })^\dag]+\operatorname{Tr}[V^{\text{ideal} }U_2^\dag +U_2(V^{\text{ideal}})^\dag]\\
    =4\cos(\varphi-\delta\phi_1 -\delta\phi_2)\Big( \cos\vartheta \cos(\delta\phi_1 -\delta\phi_2)\\ +n_z\sin\vartheta \sin (\delta\phi_1 -\delta\phi_2)\Big)\\
    +4\cos(\varphi-\delta\phi_3 -\delta\phi_4)\Big( \cos\vartheta \cos(\delta\phi_3 -\delta\phi_4)\\ +n_z\sin\vartheta \sin (\delta\phi_3 -\delta\phi_4)\Big)\, , \label{distance-1}
\end{multline}
and we are now concerned with the computation of the triple $(\varphi, \vartheta, n_z) $ maximizing the right hand side of \eqref{distance-1}. By direct computation, it is possible to prove that the maximum is attained for $$n_z=1, \, \varphi=\frac{\delta\phi_1+\delta\phi_3+\delta\phi_2+\delta\phi_4}{2},\, \vartheta= \frac{\delta\phi_1+\delta\phi_3-\delta\phi_2-\delta\phi_4}{2}\, , $$and it is equal to 
$$8\cos\left(\frac{\delta\phi_1-\delta\phi_3}{2}+\frac{\delta\phi_2-\delta\phi_4}{2}\right)\cos\left(\frac{\delta\phi_1-\delta\phi_3}{2}-\frac{\delta\phi_2-\delta\phi_4}{2}\right)$$
while the minimum square HS-distance between $U^{\text{real}}$ and $U^{\text{ideal}}$ is given by
\begin{multline}\label{baond-final-1}
    \min_{\varphi, \vartheta, \hat n}\|U^{\text{ideal}}-U^{\text{real}}\|_{HS}= \Big( 8-8\cos\left(\frac{\delta\phi_1-\delta\phi_3}{2}+\frac{\delta\phi_2-\delta\phi_4}{2}\right)\\ \cos\left(\frac{\delta\phi_1-\delta\phi_3}{2}-\frac{\delta\phi_2-\delta\phi_4}{2}\right)\Big)^{1/2}\, ,
\end{multline} 
while the operator $U^{\text{ideal}}$ is given by:
\begin{equation}
U^{\text{ideal}}=I\otimes \left(U_{\text{MMI}}U_{\text{Ph}(\phi_1,\phi_2)}V^{\text{ideal}}U_{\text{MMI}}\right)\label{Uideal}
\end{equation}
with $V^{\text{ideal}}=e^{\varphi} e^{i\vartheta \hat n \cdot \sigma}$ and with $\hat n=(0,0,1)$, $\varphi=\frac{\delta\phi_1+\delta\phi_3+\delta\phi_2+\delta\phi_4}{2}$, $ \vartheta= \frac{\delta\phi_1+\delta\phi_3-\delta\phi_2-\delta\phi_4}{2}$.
The detailed computation can be performed by considering the map $F:[0, 2\pi)\times [0, 2\pi)\times [-1, 1]\to   \mathbb{R}$ defined as
\begin{multline}
    F(\varphi, \vartheta, n_z):=4\cos(\varphi-\delta\phi_1 -\delta\phi_2)\Big( \cos\vartheta \cos(\delta\phi_1 -\delta\phi_2)\\ +n_z\sin\vartheta \sin (\delta\phi_1 -\delta\phi_2)\Big)\\
    +4\cos(\varphi-\delta\phi_3 -\delta\phi_4)\Big( \cos\vartheta \cos(\delta\phi_3 -\delta\phi_4)\\ +n_z\sin\vartheta \sin (\delta\phi_3 -\delta\phi_4)\Big)\, ,\label{function-F-gen}
\end{multline}
and compute the triple $(\varphi, \vartheta, n_z)$ for which $F$ attains its maximum.
To this end, it is convenient to represent $F$ in the equivalent form
$$F(\varphi, \vartheta, n_z)=f(\varphi)\cos\vartheta +n_z\, g(\varphi)\sin\vartheta $$
where  $f:\mathbb{R}\to \mathbb{R}$ and  $g:\mathbb{R}\to \mathbb{R}$ are given by 
\begin{equation}
    f(\varphi)=4 \Big(\cos(\varphi-\delta\phi_1 -\delta\phi_2)\cos(\delta\phi_1 -\delta\phi_2) +\cos(\varphi-\delta\phi_3 -\delta\phi_4) \cos(\delta\phi_3 -\delta\phi_4)\Big)\, ,\label{f-phi}
\end{equation}
\begin{equation}
    g(\varphi)=4 \Big(\cos(\varphi-\delta\phi_1 -\delta\phi_2)\sin(\delta\phi_1 -\delta\phi_2) +\cos(\varphi-\delta\phi_3 -\delta\phi_4) \sin(\delta\phi_3 -\delta\phi_4)\Big)\, .\label{g-phi}
\end{equation}
\subsection{A first family of local maxima}\label{subsection-appendix2a}

First of all, we observe that the partial derivative $\frac{\partial F}{\partial n_z}$ is given by
$$ \frac{\partial F}{\partial n_z}(\varphi, \vartheta, n_z)=g(\varphi)\sin\vartheta \, , $$
hence, if $(\tilde\varphi,\tilde \vartheta, \tilde n_z)$ is a point where  $\frac{\partial F}{\partial n_z}=0$, then 
\begin{enumerate}
\item[a.] either $\sin\vartheta =0$,
\item[b.] or $g(\varphi)=0$.
\end{enumerate}
\begin{enumerate}
\item[a.] In the first case case, $\cos \vartheta =\pm 1$ and the map $F $ reduces to:
$$F (\varphi, \vartheta,  n_z)=\pm f(\varphi)$$
where $f$ is given by \eqref{f-phi}.

Actually it is sufficient to find the value of $\varphi$ maximazing $ f(\varphi)$. Indeed,  by the elementary identity $\cos(\varphi +\pi)=-\cos \varphi$, it is trivial to check that both functions $f$ and $-f$ attain the same set of values.
By denoting $\varphi^*_1:=\arg \max f$, a first local maxima is attained for $\vartheta =0$ and $\varphi=\varphi^*_1$.\\

The precise value $\varphi^*_1$ can be obtained by solving the equation $f'(\varphi)=0$, with:
\begin{equation}\label{eq-phi-1}
  f'(\varphi)= - 4 \sin(\varphi-\delta\phi_1 -\delta\phi_2)\cos(\delta\phi_1 -\delta\phi_2) -4\sin(\varphi-\delta\phi_3 -\delta\phi_4) \cos(\delta\phi_3 -\delta\phi_4)=0\, .
\end{equation}
By direct computation it is easy to find all the solutions of \eqref{eq-phi-1}, which have the form $\varphi_1^*+k\pi$, $k\in \mathbb{Z} $, with  
$$\varphi_1^*=  \arctan\left(\frac{\sin(\delta\phi_1 +\delta\phi_2)\cos(\delta\phi_1 -\delta\phi_2)+\sin(\delta\phi_3 +\delta\phi_4)\cos(\delta\phi_3 -\delta\phi_4)}{\cos(\delta\phi_1 +\delta\phi_2)\cos(\delta\phi_1 -\delta\phi_2)+\cos(\delta\phi_3 +\delta\phi_4)\cos(\delta\phi_3 -\delta\phi_4)}\right).
$$
Actually,   thanks to the small values of the parameters $\delta\phi_1,\delta\phi_2,\delta\phi_3,\delta\phi_4$ ), the angle $\varphi_1^*$ satisfies the inequality $|\varphi_1^*|<\pi/2$ and gives a maximum of the function $f$ equal to:
\begin{multline}\label{M1}
    f(\varphi_1^*)=4(\cos^2(\delta\phi_1-\delta\phi_2)+\cos^2(\delta\phi_3-\delta\phi_4)\\ +2\cos(\delta\phi_1-\delta\phi_2)\cos(\delta\phi_3-\delta\phi_4)\\ \cos(\delta\phi_1+\delta\phi_2-\delta\phi_3-\delta\phi_4))^{1/2},
\end{multline}
while the angle  $\varphi_1^*+\pi $ gives a minimum.\\
\item[b.] In the second case, restricting ourselves to those value of the variable $\varphi$ such that $g(\varphi)=0$, the  function $F$ reduces to:
$$F (\varphi, \vartheta,  n_z)=f|_{g=0}(\varphi)\cos \vartheta$$
Since trivially $\max_{g=0} |f|\leq \max |f|$ and $|\cos \vartheta|\leq 1$, the possible local maxima belonging to this set of solutions do not exceed those found in step a.
\end{enumerate}

\subsection{A second family of local maxima}
Another set of points $(\varphi, \vartheta, n_z)$ maximizing locally the function \eqref{function-F-gen} can be searched among those for which $n_z=\pm 1$. In this case we are concerned with the maximization of the functions
\begin{equation}
    G_{\pm}(\varphi,\vartheta)=f(\varphi)\cos\vartheta \pm g(\varphi) \sin\vartheta\, ,
    \label{function-G-gen}
\end{equation}
wth $f$ and $g$ defined in \eqref{f-phi} and \eqref{g-phi} respectively.
Since $G_+(\varphi,2\pi -\vartheta)=G_-(\varphi,\vartheta)$, without loss of generality we can restrict ourselves to the maximization of the function $G_+$, which can be equivalently written as 
\begin{multline*}
    G_{+}(\varphi,\vartheta)=4\big(\cos(\varphi-\delta\phi_1 -\delta\phi_2)\cos(\delta\phi_1-\delta\phi_2 -\vartheta)\\ +\cos(\varphi-\delta\phi_3 -\delta\phi_4)\cos(\delta\phi_3-\delta\phi_4 -\vartheta)\big)
\end{multline*} 
The local maxima are to be searched among the stationary points of $G_+$, i.e. among the solutions of the system
\begin{equation}\left\{  \begin{array}{c}
    \frac{\partial G_+}{\partial \varphi} =0  \\
     \frac{\partial G_+}{\partial \vartheta} =0
\end{array}\right.\label{system-1}\end{equation}
More specifically:
\begin{equation}
\left\{  \begin{array}{c}
    -\sin(\varphi-\delta\phi_1 -\delta\phi_2)\cos(\delta\phi_1-\delta\phi_2 -\vartheta)\\ -\sin(\varphi-\delta\phi_3 -\delta\phi_4)\cos(\delta\phi_3-\delta\phi_4 -\vartheta) =0  \\
    \cos(\varphi-\delta\phi_1 -\delta\phi_2)\sin(\delta\phi_1-\delta\phi_2 -\vartheta)\\ +\cos(\varphi-\delta\phi_3 -\delta\phi_4)\sin(\delta\phi_3-\delta\phi_4 -\vartheta) =0
\end{array}\right.
\label{system-1.1}\end{equation}
By summing and subtracting  the two equations above, we get the equivalent system
\begin{equation}
\left\{  \begin{array}{c}
    -\sin(\varphi-2\delta\phi_1 +\vartheta) -\sin(\varphi-2\delta\phi_3 +\vartheta) =0  \\
    \sin(\varphi-2\delta\phi_2 -\vartheta) +\sin(\varphi-2\delta\phi_4 -\vartheta) =0
\end{array}\right.
\label{system-1.2}\end{equation}
 which yields the family of linear systems, labelled by two integers $k,h\in \mathbb{Z}$:
 \begin{equation}
\left\{  \begin{array}{c}
   \varphi +\vartheta =\delta\phi_1+\delta\phi_3 +k\pi \\
    \varphi -\vartheta =\delta\phi_2+\delta\phi_4 +h\pi
\end{array}\right.
\label{system-1.2}\end{equation}
with solutions
\begin{equation}
\left\{  \begin{array}{c}
    \varphi=\frac{\delta\phi_1+\delta\phi_3+\delta\phi_2+\delta\phi_4}{2}+\frac{k+h}{2}\pi\, ,\\
    \vartheta= \frac{\delta\phi_1+\delta\phi_3-\delta\phi_2-\delta\phi_4}{2}+\frac{k-h}{2}\pi\, .
\end{array}\right.
\end{equation}
Under the assumption that the fluctuations $\delta\phi_1,\delta\phi_3,\delta\phi_2,\delta\phi_4$ are small, the maximum of the map $G_+$ is attained at $\varphi=\frac{\delta\phi_1+\delta\phi_3+\delta\phi_2+\delta\phi_4}{2},\vartheta= \frac{\delta\phi_1+\delta\phi_3-\delta\phi_2-\delta\phi_4}{2} $ and it equals to
\begin{multline}\label{M2}
    G_+ \left(\frac{\delta\phi_1+\delta\phi_3+\delta\phi_2+\delta\phi_4}{2},\frac{\delta\phi_1+\delta\phi_3-\delta\phi_2-\delta\phi_4}{2}\right) \\ =8\cos\left(\frac{\delta\phi_1-\delta\phi_3}{2}+\frac{\delta\phi_2-\delta\phi_4}{2}\right)\cos\left(\frac{\delta\phi_1-\delta\phi_3}{2}-\frac{\delta\phi_2-\delta\phi_4}{2}\right)
\end{multline}

If the fluctuations $\delta\phi_1,\delta\phi_3,\delta\phi_2,\delta\phi_4$ are small, one can verify that the local maximum \eqref{M2} is greater than the local maximum \eqref{M1}. Hence the absolute maximum is attained for 
$$n_z=1, \, \varphi=\frac{\delta\phi_1+\delta\phi_3+\delta\phi_2+\delta\phi_4}{2},\, \vartheta= \frac{\delta\phi_1+\delta\phi_3-\delta\phi_2-\delta\phi_4}{2}\, .$$
 \subsection{Appendix: Analytical bound on the HS norm}
We can now introduce a rather conservative bound by computing the maximum of \eqref{baond-final-1} over the admissible range of the phase fluctuations. In particular, by assuming that 
$$|\delta\zeta_1|\leq \epsilon, \, |\delta\zeta_2|\leq \epsilon\, ,|\delta\zeta_3|\leq \epsilon\,, |\delta\zeta_4|\leq \epsilon, $$
for a suitable constant $\epsilon>0$, then 
\begin{equation}\label{baond-final-2}
\max_{\delta\zeta_1,\delta\zeta_2,\delta\zeta_3,\delta\zeta_4}\, \min_{\varphi, \vartheta, \hat n}\|U^{\text{ideal}}-U^{\text{real}}\|_{HS}
\leq 2\sqrt 2\left( 1-\cos\left(2\epsilon \right)\right)^{1/2}.
\end{equation} 
In particular, by expanding the r.h.s of \eqref{baond-final-2} in powers of $\epsilon$ and by neglecting the terms $\epsilon ^k $ of order $k\geq 2$, we eventually get:
\begin{equation}\label{bound-final-3}
\max_{\delta\zeta_1,\delta\zeta_2,\delta\zeta_3,\delta\zeta_4}\, \min_{\varphi, \vartheta, \hat n}\|U^{\text{ideal}}-U^{\text{real}}\|_{HS}\leq 4 \epsilon. 
\end{equation}
Considering now the two operators $$U_{\text{real}}(\phi_1,\phi_2,\delta\phi_1,\delta\phi_2,\delta\phi_3,\delta\phi_4)=U^{\text{real}}_\phi$$ and $$U_{\text{real}}(\theta_1,\theta_2,\delta\theta_1,\delta\theta_2,\delta\theta_3,\delta\theta_4)=U^{\text{real}}_\theta.$$
Similarly, the factorized operator minimizing the distance from the $U^{\text{real}}_\phi$ (resp. $U^{\text{real}}_\theta$) will be denoted $U^{\text{ideal}}_\phi$ (resp. $U^{\text{ideal}}_\theta$ ).
The unitary operator describing the overall action of the rotation stage is given by the product $U^{\text{real}}_\phi U^{\text{real}}_\theta$ and its HS-distance from the product of the two ideal (factorized) operators $U^{\text{ideal}}_\phi U^{\text{ideal}}_\theta$ can be estimated as
\begin{multline}
    \|U^{\text{real}}_\phi U^{\text{real}}_\theta-U^{\text{ideal}}_\phi U^{\text{ideal}}_\theta\|_{HS}\\
    \leq \|(U^{\text{real}}_\phi-U^{\text{ideal}}_\phi)(U^{\text{real}}_\theta-U^{\text{ideal}}_\theta)\|_{HS} \\
    +\|U^{\text{ideal}}_\theta (U^{\text{real}}_\phi-U^{\text{ideal}}_\phi)\|_{HS}+
   \| (U^{\text{real}}_\theta-U^{\text{ideal}}_\theta)U^{\text{ideal}}_\phi\|_{HS}\\
   \leq \|(U^{\text{real}}_\phi-U^{\text{ideal}}_\phi)\| \|(U^{\text{real}}_\theta-U^{\text{ideal}}_\theta)\|_{HS} \\
   +\|U^{\text{ideal}}_\theta \| \|(U^{\text{real}}_\phi-U^{\text{ideal}}_\phi)\|_{HS}+
   \| (U^{\text{real}}_\theta-U^{\text{ideal}}_\theta)\|_{HS} \|U^{\text{ideal}}_\phi\|\\
   \leq \|(U^{\text{real}}_\phi-U^{\text{ideal}}_\phi)\|_{HS} \|(U^{\text{real}}_\theta-U^{\text{ideal}}_\theta)\|_{HS} \\
   + \|(U^{\text{real}}_\phi-U^{\text{ideal}}_\phi)\|_{HS}+
   \| (U^{\text{real}}_\theta-U^{\text{ideal}}_\theta)\|_{HS} 
\end{multline}
In particular, by assuming that the phase fluctuations have maximum amplitude $\epsilon$ and by neglecting the terms  $\epsilon ^k$ of order $k\geq 2$, we get the final estimate:
\begin{equation}\label{bound-final-4}
\max_{\delta\phi_i,\delta\theta_j}\, \min_{\varphi, \vartheta, \hat n}\|U^{\text{ideal}}-U^{\text{real}}\|_{HS}\leq 8\sqrt 2 \epsilon 
\end{equation}

\section{Appendix: Entropy certification based on Bell inequality violation}\label{appendix-theoreticalpreliminaries}
For clarity, in the following discussion we identify the guessing probability of the main text $\mathbb{P}_{\text{guess}}$ with $G$ to better distinguish it from the probabilities indicated as $\mathbb{P}$.
The relevant figure of merit of a device-independent certification protocol based on CHSH violation is the (realization-independent) {\em quantum  guessing probability}  $G(\mathbb{P})$ associated to a (quantum) probability distribution $\{\mathbb{P}(x,y)\}_{x,y=\pm 1}$ defined as
  \begin{equation}
G(\mathbb{P}):=
\sup_{ \{\tilde \rho, A, B\} \in R( \mathbb{P} ) }    G(\tilde \rho, A, B), \label{real-Gp}
\end{equation}
where  the family $R(\mathbb{P})$ contains 
all possible realizations $\{\tilde \rho, A,B\}$ compatible with $\mathbb{P}$, i.e. all pairs of quantum states $\tilde \rho$ and product observables $A\otimes B$, each with two possible outcomes $x,y\in \{\pm 1 \}$, such that
$$\mathbb{P}(x,y)=\operatorname{Tr}\left[\tilde \rho P_x^A\otimes P_y^B\right]\,\quad x,y=\pm 1\,,$$
$\{P_x^A\}_{x=\pm 1}$ and $\{P_y^B\}_{y=\pm 1}$ being the PVMs associated to $A$ and $B$ respectively.
The corresponding min entropy is defined as $H_{\text{min}}:=-\log_2(G(\mathbb{P}))$.

For a generic mixed state $\rho$ the {\em average guessing probability} $G(\tilde \rho, A, B)$ appearing on the r.h.s of \eqref{real-Gp} is defined as
\begin{equation}G(\tilde \rho, A, B)=\sup\int  G(\psi_\lambda, A,B)d\nu (\lambda)\label{Gp-mixed}\end{equation}
where the supremum is taken over all decompositions $\rho=\int |\psi_\lambda\rangle \langle\psi_\lambda|d\nu (\lambda)$ of $\rho$ into an incoherent superposition of pure states $|\psi_\lambda\rangle$ ($\nu$ being a probability measure), 
while 
\begin{equation}
G(\psi_\lambda, A,B):=\max_{(x,y)}\operatorname{Tr}\left[|\psi_\lambda\rangle \langle\psi_\lambda|P_x^A\otimes P_y^B\right]\,
.\label{Gp-pure}
\end{equation}
Giving a pure state $|\psi\rangle $ and two pairs of observables $A_1,A_2$, and $B_1,B_2$ yielding a value $\chi$ for the correlation function
\begin{equation}
    \chi=<A_1\otimes B_1>_{\psi}+<A_1\otimes B_2>_{\psi} +<A_2\otimes B_1>_{\psi}-<A_2\otimes B_2>_{\psi}
\end{equation}
the  inequality $G(\psi_\lambda, A_i,B_j)\leq f(\chi)$, with $f(x)=\frac{1}{2}+\frac{1}{2}\sqrt{2-\frac{x^2}{4}}$,  holds true for any $i,j=1,2$ \cite{Pironio10,Acin12}. By the concavity of the function $f$, this inequality can be generalized to the case of convex superpositions of pure states of the form $\rho=\int |\psi_\lambda\rangle \langle\psi_\lambda|d\nu (\lambda) $, even considering the case of two pairs of observables $A(\lambda)_1,A(\lambda)_2$, and $B(\lambda)_1,B(\lambda)_2$ explicitly depending on the parameter $\lambda$ and yielding a CHSH parameter $\chi_\lambda$
\begin{equation}
    \chi_\lambda=<A^\lambda_1\otimes B^\lambda_1>_{\psi_\lambda}+<A^\lambda_1\otimes B^\lambda_2>_{\psi_\lambda}\\ +<A^\lambda_2\otimes B^\lambda_1>_{\psi_\lambda}-<A^\lambda_2\otimes B^\lambda_2>_{\psi_\lambda}
\end{equation}

In particular:

\begin{align}
&\max_{(x,y)}\int \operatorname{Tr}\left[|\psi_\lambda\rangle \langle\psi_\lambda|P_x^{A(\lambda)_i}\otimes P_y^{B(\lambda)_j}\right]d\nu (\lambda)\\
\leq &\int \max_{(x,y)} \operatorname{Tr}\left[|\psi_\lambda\rangle \langle\psi_\lambda|P_x^{A(\lambda)_i}\otimes P_y^{B(\lambda)_j}\right]d\nu (\lambda)\\
\leq &\int f(\chi_\lambda)d\nu(\lambda)\leq f\left(\int \chi_\lambda d\nu(\lambda)\right)=f(\chi)\label{bound-Bell}
\end{align}

The chain of inequalities above, the concavity of the function f, and the explicit dependence of the final term in \eqref{bound-Bell} only on the parameter $\chi$, i.e. a function of the probability distributions $\{\mathbb{P}^{i,j}(x,y)\}_{x,y=\pm 1}$ independent of their particular quantum realizations, allow to prove the final bound for the quantum guessing probability of each of the four  distributions $\{\mathbb{P}^{i,j}(x,y)\}_{x,y=\pm 1}$:
\begin{equation}
    \label{final-boundQGP}
    G(\mathbb{P}^{i,j})\leq f(\chi)\qquad \forall i,j=1,2\,.
\end{equation}
Inequality \eqref{final-boundQGP} is actually robust under a generalization of definition \eqref{Gp-mixed}. Indeed, thanks to the discussion leading to \eqref{bound-Bell}, for each quantum probability distribution $\{\mathbb{P}(x,y)\}_{x,y=\pm 1}$ one can consider a larger set of realizations allowing, at least in principle, to change observables according to the different components of a mixed state. More precisely, given a quantum state $\rho$ and any decomposition $\rho=\int |\psi_\lambda\rangle \langle\psi_\lambda|d\nu (\lambda)$, one can considers corresponding product observables $A(\lambda) \otimes B(\lambda)$ such that 
$$\mathbb{P}(x,y)= \int \operatorname{Tr}\left[|\psi_\lambda\rangle \langle\psi_\lambda|P_x^{A(\lambda)}\otimes P_y^{B(\lambda)}\right].$$

\end{backmatter}

\bibliography{sample}






\end{document}